\documentclass[prd,onecolumn]{revtex4}
\usepackage{dcolumn}
\usepackage{multirow}
\usepackage{graphicx}
\usepackage{amssymb}
\usepackage{bm}
\usepackage{hyperref}
\usepackage{epstopdf}
\usepackage{color}
\usepackage{mathrsfs}
\usepackage{amsmath,amssymb,amsthm}
\usepackage{rotating}
\usepackage{sverb, longtable}
\usepackage{subfigure}

\usepackage[graphicx]{realboxes}
\usepackage{adjustbox}
\begin{document}
\title{Black Hole Olympics: Metric Selection from the Event Horizon Telescope and LIGO-VIRGO Observations}
\author{Deng Wang}
\email{cstar@nao.cas.cn}
\affiliation{National Astronomical Observatories, Chinese Academy of Sciences, Beijing, 100012, China}
\begin{abstract}
We propose a new paradigm in the field of black hole physics, i.e., with more and more high precision observations, one must make a metric selection to determine which black hole is observationally preferred. In light of the shadow imaging data from the Event Horizon Telescope and gravitational wave measurements from the LIGO-VIRGO collaboration, we attempt to address this issue.
Although not finding any obvious preference for a specific black hole metric based on current data, as a presentation of this new research paradigm, we rank these black hole metrics using the Bayesian information criterion. We find that Kerr and Reissner-Nordstr\"{o}m tie the first place in the black hole Olympics Games. Interestingly, we give the $2\,\sigma$ upper bound of the average electronic charge $Q<2.82\times10^{18}$ C for the Reissner-Nordstr\"{o}m spacetime, and the $2\,\sigma$ constraint on the average spin parameter $a=-0.02^{+4.07}_{-4.04}$ m for the Kerr spacetime, which is consistent with the prediction of zero rotation for a distant observer. Future multi-messenger and multi-wavelength observations will enhance the application of this new paradigm into tests of gravitational theories and black holes.   	
\end{abstract}
\maketitle

\section{Introduction}
Since Einstein firstly proposed general relativity (GR) in 1915 \cite{Einstein1915}, GR has been tested in light of various kinds of observations ranging from large scales to very small scales \cite{Will:2014kxa}. Cosmological observations leading to the discovery of the late-time accelerated expansion give tight constraints on possible deviations from GR at both the background and perturbation levels \cite{Ferreira:2019xrr}. At the radio wavelengths, pulsar observations \cite{Stairs:2003eg,Kramer:2021jcw,Wex:2020ald} in the binary systems provide a good platform to test the matter-gravity coupling and investigate the radiative properties predicted by GR in the strong-field regions. Within our solar system, observations such as the gravitational time delays when photons pass the solar surface \cite{Bertotti:2003rm}, the deflection of a light ray when solar eclipses occur \cite{Lambert:2011} and the precession of Mercury's orbit \cite{Verma:2013ata}, have verified the validity of GR to a high precision.     

Even though GR is successful based on current observations \cite{Baker:2014zba}, there still exists many intriguing puzzles within its framework such as: (i) Is actually GR correct on cosmological scales at all? (ii) Is the dynamics of early universe governed by GR? Especially, it is urgent to answer the question that whether GR is valid at horizon scales when being applied into black hole systems. Since a mature gravity theory has unique effects on the phenomena of black holes, one can theoretically study the properties of a gravity theory by analyzing its corresponding black hole solutions \cite{Harlow:2014yka,Wang:2022ivi,Vagnozzi:2022moj}. As a consequence, black holes have become an excellent platform to probe the gravity theory. However, an important problem that which black hole solution is observationally preferred emerges. To address this issue, high precision observations are needed. 
As is well known, in astrophysical observations, although black holes have been found in a wide range of masses including supermassive ones \cite{Webster1972,Remillard:2006fc} and stellar-mass ones \cite{Lynden-Bell:1969gsv,Kormendy:1995er,Miyoshi:1995da}, we still lack the phenomenological knowledge of their event horizons in observational aspects over a long period of time. 
Until the recent several years, nonetheless, observations of black holes become realistic. This fact allows us to investigate the black hole physics and probe the nature of gravity better.
More and more mature gravitational wave observations from the LIGO-VIRGO collaboration \cite{LIGOScientific:2016aoc,LIGOScientific:2016lio,LIGOScientific:2019fpa,LIGOScientific:2020tif}, the detection of relativistic effects in the orbits of stars around Sgr A* \cite{GRAVITY:2018ofz,GRAVITY:2020gka,Do:2019txf},  
and the imaging data of M87* shadow from the Event Horizon Telescope (EHT) \cite{EventHorizonTelescope:2019dse,EventHorizonTelescope:2019ggy,EventHorizonTelescope:2019pgp,EventHorizonTelescope:2019ths}, which is a global very long baseline interferometry working at a wavelength of 1.3 mm (230 GHz), permit us to explore the physical processes and spacetime geometry around the event horizon of a black hole. Particularly, the EHT imaging survey provides an excellent opportunity for us to study the dark shadow induced by gravitational light bending and photon capture at the horizon scale.   

Recently, the EHT collaboration \cite{EventHorizonTelescope:2022xnr,EventHorizonTelescope:2022wok,EventHorizonTelescope:2022xqj} has released the first image of Sgr A* in the Galactic center. This horizon-scale image can help test the black hole physics and GR predictions better than the image of M87*. Since the mass and distance of Sgr A* is accurately known by the detection of relativistic effects in stellar orbits \cite{Do:2019txf}, strong constraints can be implemented on the spacetime properties. In this study, for the first time, we identify the emergent problem that which black hole solution is the best at all, and attempt to implement the first black hole metric selection by combining the shadow measurements of Sgr A* and M87* with gravitational wave observations \cite{EventHorizonTelescope:2019dse,EventHorizonTelescope:2022xnr}. We find that there is no obvious preference for a specific black hole metric based on current observations, and give the constraints on the electronic charge in the Reissner-Nordstr\"{o}m (RN) solution and the angular momentum in the Kerr spacetime. 

This study is outlined in the following manner. In Section II, we describe the black hole metrics to be constrained. In Section III, we briefly review the approach to calculate the diameter of the shadow cast by a black hole. In Section IV, we introduce the data and analysis methodology and present the numerical results. In the final section, we discuss and conclude.

\section{Black hole metrics}
In order to determine which black hole metric is preferred via observations, in principle, one can test any black hole solutions in any extension to GR. Furthermore, the nature of gravity can also be tested. In this analysis, we just propose the new paradigm of comparing different black hole metrics by using current data. As a concrete example, we will focus on classical black hole spacetime configurations in GR. For the sake of simplicity, we consider a parameterized Kerr-Newman metric \cite{Wald:1984} whose line element in Boyer-Lindquist coordinates reads as  
\begin{equation}
ds^2=-\left(1-\frac{2\alpha Mr}{\rho^2}\right)dt^2+\frac{\rho^2}{\Delta}dr^2+\rho^2d\theta^2-\frac{4a\alpha Mr\mathrm{sin^2}\theta}{\rho^2}dtd\phi+\frac{\left[(a^2+r^2)^2-a^2\Delta\mathrm{sin^2}\theta\right]\mathrm{sin^2}\theta}{\rho^2}d\phi^2,       \label{1}
\end{equation}
where the metric functions are 
\begin{equation}
\Delta\equiv r^2-2\alpha Mr+a^2+Q^2, \quad \rho^2\equiv r^2+a^2\mathrm{cos^2}\theta,  \label{2}
\end{equation}
where $M$, $Q$ and $a$ ($a\equiv J/M$) are the mass, electronic charge and reduced angular momentum of the black hole, respectively, and the parameter $\alpha$ characterizes the possible deviation from Kerr-Newman spacetime. Note that here we use the units $G=c=4\pi\epsilon_0=1$, where $G$, $c$ and $\epsilon_0$ denote the Newton's gravitational constant, speed of light, and vacuum permittivity, respectively. It is easy to find that this metric reduces to the Schwarzschild solution when $Q=a=0$ and $\alpha=1$. 

To perform the model selection of black holes, we take the following four metrics: (i) Schwarzschild; (ii) RN; (iii) Kerr; (iv) parameterized Schwarzschild ($Q=a=0$). It is worth noting that in our numerical analysis, the Schwarzschild black hole has one free parameter and the other three have two parameters. 

\section{Black hole shadows}
For a general static spherically symmetric black hole, its line element is written as
\begin{equation}
ds^2 = -g_{tt}dt^2+g_{rr}dr^2+r^2d\theta^2+r^2\mathrm{sin^2}\theta d\phi^2, \label{3}
\end{equation}  
where $g_{tt}$ and $g_{rr}$ characterize the time and radial components of the spacetime, respectively. Following Ref.\cite{EventHorizonTelescope:2020qrl} and using two of the Killing vectors, one can express the components of the momentum of a photon flying in the spacetime as
\begin{equation}
(k^t,\,k^r,\,k^\theta,\,k^\phi)=\left(\frac{E}{g_{tt}},\,\sqrt{-\frac{E^2}{g_{tt}g_{rr}}-\frac{l^2}{g_{rr}r^2}},\,0,\,
\frac{l}{r^2}\right), \label{4}
\end{equation} 
where $E$ and $l$ denote the energy and angular momentum of a photon. Notice that we have fixed $\theta=\pi/2$ to implement the analysis. The radius $r_{ps}$ of the photon sphere of a black hole can be determined by the following two conditions
\begin{equation}
k^r=0, \quad \frac{dk^r}{dr}=0. \label{5}
\end{equation}
Using Eq.(\ref{5}), $r_{ps}$ is easily obtained by solving the equation
\begin{equation}
2g_{tt}-r\frac{dg_{tt}}{dr}=0. \label{6}
\end{equation} 
It is well known that a black hole shadow is a gravitationally lensed image of the circular photon orbit. After considering this effect, the shadow diameter is shown as
\begin{equation}
d_{sh}=\frac{2r_{ps}}{g_{tt}(r_{ps})}.  \label{7}
\end{equation}
This quantity will be used as our theoretical input in the numerical analysis.

\section{Data, methodology and results}
As mentioned above, we will constrain black hole geometries using the data combination of EHT shadow and LIGO-VIRGO observations. For Sgr A* from EHT, we take its angular shadow diameter $48.7\pm7$ $\mu$as \cite{EventHorizonTelescope:2022xnr} and the estimated mass from VLTI stellar orbit monitoring observations $4.297\pm0.013\times10^6$ M$_\odot$ \cite{Do:2019txf}. For M87* from EHT, we derive its angular shadow size using the measured angular gravitational radius $3.8\pm0.4$ $\mu$as \cite{EventHorizonTelescope:2019dse} and its fractional deviation $-0.01\pm0.17$ from the Schwarzschild solution \cite{EventHorizonTelescope:2019ggy}. In Fig.25 in Ref.\cite{EventHorizonTelescope:2022xqj}, since the EHT collaboration gives the posterior distributions of fractional deviations from the Schwarzschild predictions for two low-mass black hole binary mergers GW190924 and GW170608, we use these two distributions to derive the ``effective'' shadow diameters as extra data points. Specifically, first of all, we derive the corresponding 1 $\sigma$ errors of deviations from the Schwarzschild case, and then derive the angular gravitational radius using their measured masses \cite{LIGOScientific:2020kqk}. Finally, similar to M87*, we obtain their shadow diameters using the gravitational radii and fractional deviations from the Schwarzschild predictions. In total, we have four shadow diameter points to implement the constraints.

In principle, black holes with different masses can have different electric charges and spins. By performing the MCMC analysis, one can obtain different upper bounds on $Q$ and $a$ for different black holes. However, if we want to know the average $Q$ and $a$ for a large population of black holes, we should combine all black hole shadow observations together to implement the constraints.   

To confront black hole metrics with observations, we use the standard $\chi^2$ statistics        
\begin{equation}
\chi^2(\delta,\,p)=\sum\limits_{i=1}^n\frac{\left[\delta-\delta_{obs}(p)\right]^2}{\sigma_{obs}^2},  \label{8}
\end{equation}
where $\delta_{obs}$ is defined as
\begin{equation}
\delta_{obs}\equiv\frac{d_{sh,obs}}{d_{sh,th}(p)}-1,  \label{9}
\end{equation}

\begin{figure}
	\centering
	\includegraphics[scale=0.5]{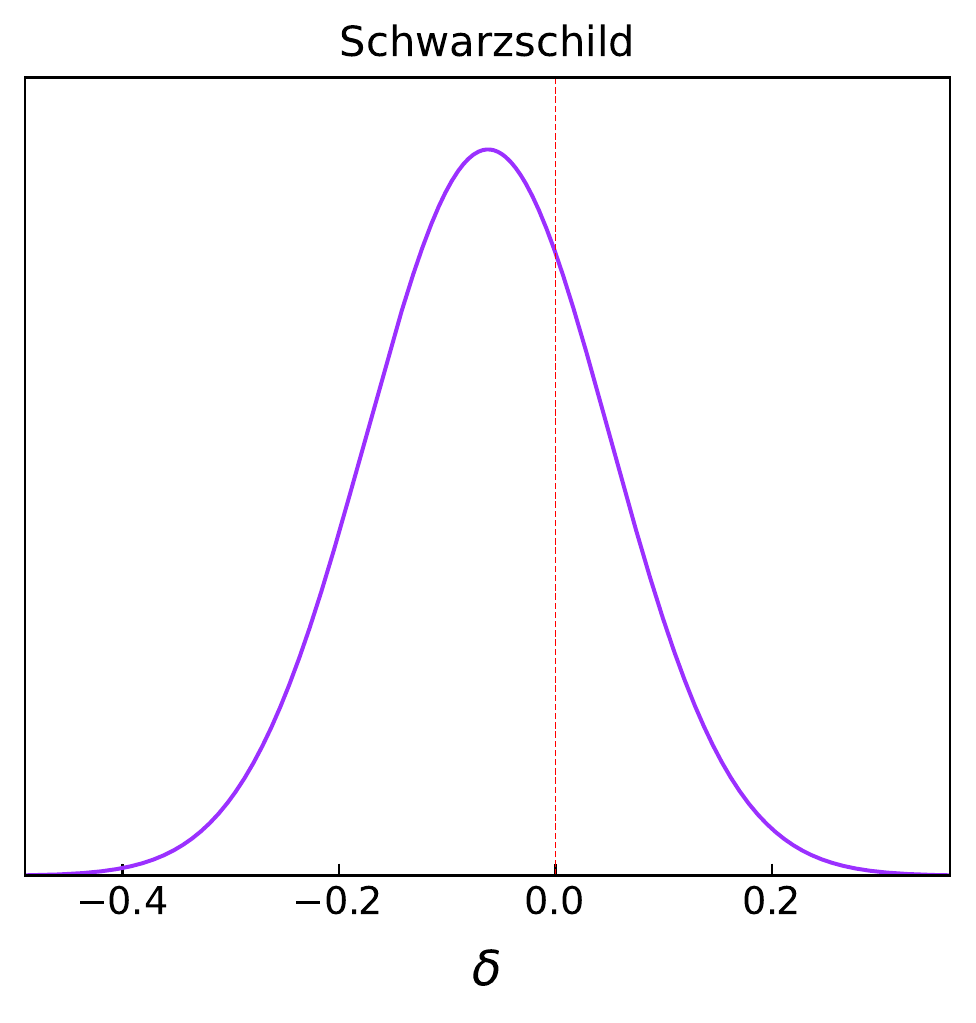}
	\caption{The 1-dimensional normalized posterior distribution of the fractional deviation from the Schwarzschild spacetime. The vertical line denotes the Schwarzschild case.}\label{f1}
\end{figure}

\begin{figure}
	\centering
	\includegraphics[scale=0.5]{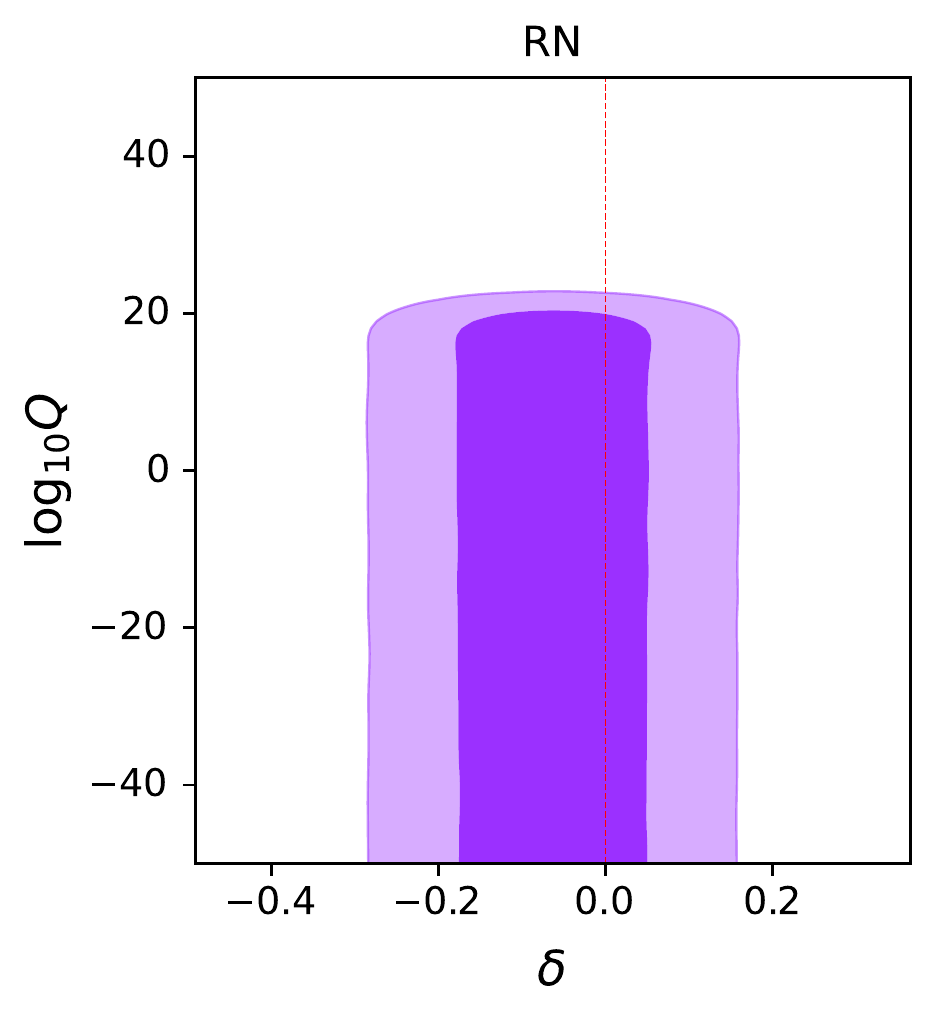}
    \includegraphics[scale=0.5]{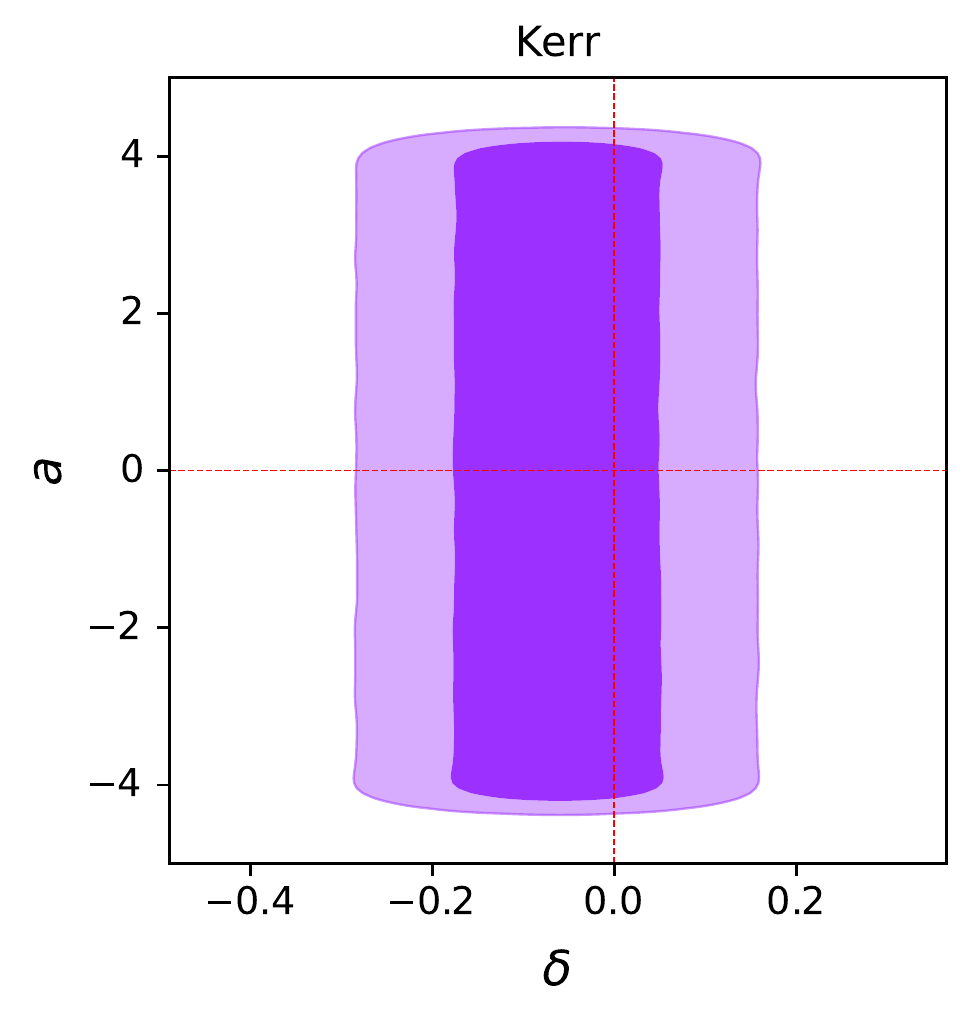}
    \includegraphics[scale=0.5]{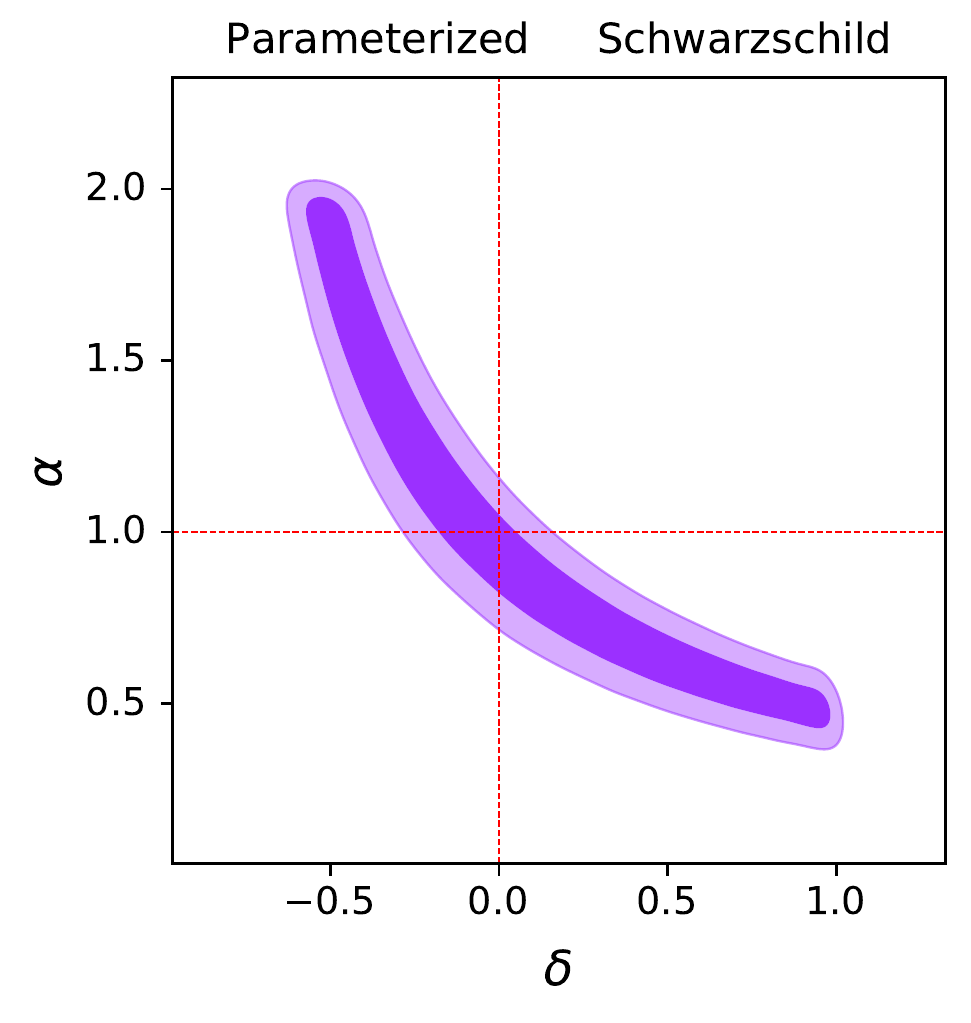}
	\caption{The 2-dimensional posterior distributions of typical parameters are shown for the RN, Kerr and parameterized Schwarzschild black hole solutions, respectively. The cross points of horizontal and vertical lines denote the the Schwarzschild case. Specially, the horizontal line lies at the negative infinity for RN.}\label{f2}
\end{figure}

\begin{table}[h]
	\renewcommand\arraystretch{1.5}
	\caption{The constraints on the Schwarzschild, RN, Kerr and parameterized Schwarzschild black hole metrics are shown. We also announce the rank in the Black Hole Olympics.}
	\setlength{\tabcolsep}{5mm}{
		{\begin{tabular}{@{}ccccc@{}} \toprule
				Parameters            &Schwarzschild           &RN      &Kerr           &Parameterized Schwarzschild                              \\ \colrule
				$\delta$       &-0.063$\pm$0.109  &-0.063$\pm$0.109  &$-0.063^{+0.215}_{-0.216}$   &$-0.037^{+0.619}_{-0.376}$           \\  
				$\mathrm{log_{10}}\,Q$       &---      &$<18.45$ $(2\,\sigma)$    &---                    &---                                              \\
				$a$           &---              &---           &$-0.02^{+4.07}_{-4.04}$ $(2\,\sigma)$            &---                           \\
				$\alpha$          &---            &---         &---            &$0.97^{+0.62}_{-0.38}$                           \\
				\hline
				BIC &1.171 &2.557 &2.557  &2.550 \\
				\hline 
				Rank  &\begin{minipage}[b]{0.1\columnwidth}
					\raisebox{-.5\height}{\includegraphics[scale=0.05]{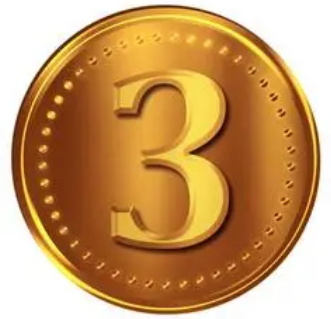}}
				\end{minipage}  
				&\begin{minipage}[b]{0.1\columnwidth}
					\raisebox{-.5\height}{\includegraphics[scale=0.05]{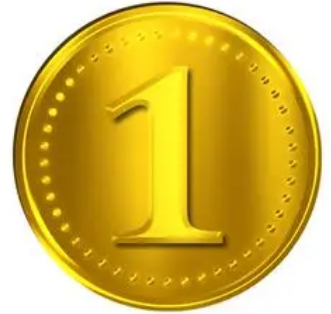}} 
				\end{minipage}  
				&\begin{minipage}[b]{0.1\columnwidth}
					\raisebox{-.5\height}{\includegraphics[scale=0.05]{gold.png}} 
				\end{minipage}
				&\begin{minipage}[b]{0.1\columnwidth}
					\raisebox{-.5\height}{\includegraphics[scale=0.05]{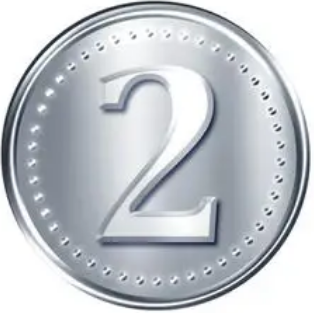}} 
				\end{minipage}        \\
				\botrule
			\end{tabular}
			\label{t1}}}
\end{table}

where $d_{sh,th}$, $d_{sh,obs}$ and $\sigma_{obs}$ are the theoretical shadow diameter, observed shadow size and uncertainties of observations, respectively. $n$ is the number of data points, and $\delta$ and $p$ are free parameters which characterize the fractional deviation from the corresponding spacetime and the black hole metric considered.

Furthermore, we take the Bayesian analysis and Markov Chain Monte Carlo \cite{ForemanMackey:2012ig} to obtain the posterior distributions of parameters. When implementing the calculations, we use the SI units. To analyze the chains, we take the public package \texttt{Getdist} \cite{Lewis:2019xzd}. Our numerical results are presented in Figs.\ref{f1}-\ref{f2} and Tab.\ref{t1}. Combining the EHT with gravitational wave observations, we obtain the constraint $\delta=-0.063\pm0.109$ on the deviation from Schwarzschild, which is very consistent with the result from the EHT collaboration \cite{EventHorizonTelescope:2019ggy}. In Fig.\ref{f1}, one can easily find that this constraint prefers slightly a negative deviation, which may indicate that for a distant observer, observations support a smaller shadow size than the Schwarzschild metric predicts. Furthermore, we study the deviation from RN by using four data points, and obtain $\delta=-0.063\pm0.109$, which is the same as the Schwarzschild case and implies that the electronic charge has an extremely small effect on the shadow size from the viewpoint of observations. Interestingly, we give the $2\,\sigma$ upper bound of the average electronic charge $Q<2.82\times10^{18}$ C, which is looser than the astrophysical constraint $Q\lesssim3\times10^8$ C \cite{Zajacek:2018ycb}. Subsequently, we investigate the deviation from the important Kerr spacetime and give the constraint $\delta=-0.063^{+0.215}_{-0.216}$. Although the mean value remains, this deviation has a larger error than the above two solutions by a factor of $\sim$ 2. In the meanwhile, we give the $2\,\sigma$ constraint on the average spin parameter $a=-0.02^{+4.07}_{-4.04}$ m, which is compatible with the prediction of zero rotation for a distant observer. Moreover, for the parameterized Schwarzschild spacetime, we find $\delta=-0.037^{+0.619}_{-0.376}$ and $\alpha=0.97^{+0.62}_{-0.38}$ at the $1\,\sigma$ confidence level. Even though this one-parameter metric can characterize the possible deviation from Schwarzschild for a black hole configuration, our constraint can only provide relatively loose constraint on two parameters. From Fig.\ref{f2}, it is easy to see that these two parameters are strongly anti-correlated. The reason is that, in Eq.\ref{9}, $\delta$ is inversely proportional to $\alpha$. Actually, the data can give a very tight constraint on the compound parameter $\beta\equiv\alpha\delta$. Since our goal is amongst the first to compare clearly different black hole metrics in light of observations, we just adopt the parameterized Schwarzschild metric as an example.         

In order to make a comparison between different metrics, we employ the Bayesian information criterion (BIC) \cite{Schwarz} as our selection criterion. This criterion favors the model which have fewer parameters when giving the same fit to data. The BIC is usually defined as BIC $=\chi^2_{min}+k\,\mathrm{ln}\,n$, where $k$ denotes the number of free parameters for a model. After simple calculations, we find the BIC values are, respectively, 1.171, 2.557, 2.557 and 2.550 for Schwarzschild, RN, Kerr and parameterized Schwarzschild solutions. Because the differences $\Delta$ BIC$\sim$2 and $\Delta$ BIC$\sim$ 6 indicate a positive and strong evidence against the reference model, we can not conclude any obvious preference for a specific black hole metric considered here based on current data. However, as a presentation, we can still use the BIC values to rank these four black holes. After a severe competition, the RN and Kerr black holes win the championship, parameterized Schwarzschild solution takes the silver medal, and the simplest Schwarzschild spacetime win the bronze.     

\section{Discussions and conclusions}
For a long time, in the research of gravitational theories, one can theoretically study the specific black hole solutions without obvious limitations. However, the recent black hole imaging survey by the EHT collaboration poses a direct challenge to the traditional black hole research. Besides M87*, especially, the release of Sgr A* image together with high-precision stellar orbit measurements from VLTI gives an important restriction to the underlying black hole spacetime geometries. This means that from now on, a black hole researcher must confront obtained black hole solutions with shadow data, in order to ensure the validity of ideas. As a consequence, an important topic that which black hole is the best emerges. 

To address this issue, we constrain four black hole metrics and make a simple model selection among them by using two shadow measurements from M87* and Sgr A* and two effective shadow data points from LIGO-VIRGO's gravitational wave observations. 

We find that there is no evidence for any deviation from all four metrics, since the values of $\delta$ are all consistent with zero at the $1\,\sigma$ confidence level. Furthermore, we give the $2\,\sigma$ upper bound of the average electronic charge $Q<2.82\times10^{18}$ C in the RN case, and the $2\,\sigma$ constraint on the average angular momentum parameter $a=-0.02^{+4.07}_{-4.04}$ m in the Kerr case, which is compatible with the prediction of zero rotation for a distant observer.
Interestingly, via the BIC, we hold the Olympic Games for four black holes, which act as the representatives of all black hole candidates. After a fierce competition, the RN and Kerr black holes win the championship, parameterized Schwarzschild solution takes the silver medal, and the simplest Schwarzschild spacetime gets the bronze. 

It should be stressed that for a large population of black holes across many mass scales, to determine which black hole spacetime is observationally preferred, our method needs to assume the global average charge and spin in order to carry out the constraints.

It is obvious that the research of black hole physics will enter the multi-messenger era, by combining the EHT with other observations such as gravitational wave measurements from the LIGO-VIRGO collaboration. Future multi-wavelength and multi-messenger data will push the the research of gravitational theories and black holes to a new stage.  

\section{Acknowledgments}
DW thank Changjun Gao and Yuan Sun for helpful discussions. This work is supported by the National Science Foundation of China under Grants No.11988101 and No.11851301.

\end{document}